\documentclass{elsart}
\usepackage{epsfig}

\usepackage{natbib}
\begin{document}
\runauthor{Vestrand and Sreekumar}
\begin{frontmatter}
\title{RXTE Observations of PKS 2155-304 during the November 1997 Gamma-Ray Outburst}
\author[address 1]{W. Thomas Vestrand} and
\author[address 2]{P. Sreekumar}


\address[address 1]{Space Science Center,
 University of New Hampshire, Durham, NH 03824 }
\address[address 2]{Laboratory for High Energy Astrophysics, NASA/GSFC, Greenbelt, MD 20771}

\begin{abstract}
We present x-ray observations of the nearby BL Lac PKS 2155-304 taken when it was
undergoing a GeV/TeV gamma-ray outburst.  During the outburst we measured x-ray fluxes
in the 2-10 keV band that are the largest ever observed for PKS 2155-304.  Comparison
of these November 1997 measurements and other x-ray observations made contemporaneously
with GeV or TeV gamma-ray observations indicate that x-ray and gamma-ray emissions 
are correlated. Measurements with x-ray all-sky monitors such as the ASM/RXTE and MOXE can therefore signal the presence of
outbursts at gamma-ray energies from PKS 2155-304.
\end{abstract}
\end{frontmatter}


\typeout{SET RUN AUTHOR to \@runauthor}

\section{Introduction}

PKS 2155-304 is the archetypical X-ray-selected BL Lac object (XBL). It is one of the
brightest BL Lacs at x-ray through optical wavelengths where it has a relatively featureless
continuum and displays rapid, large amplitude variability. This continuum is thought 
to be direct synchrotron emission from a distribution of ultra-relativistic electrons which
extends to unusually high energies \cite{E1} .

The gamma-ray emission from PKS 2155-304 constitutes a second, separate, spectral component.
Observations with the EGRET telescope aboard the Compton Gamma Ray Observatory (CGRO) show that the
spectral energy distribution of this gamma-ray component must peak at energies above 10 GeV \cite{V1}. This,
plus the realization that the extention of the synchrotron component into the x-ray band meant that
ambient photons would be scattered to TeV energies, led to predictions \cite{V1,S1} that PKS 2155-304 would be a
detectable  TeV gamma ray source.

The University of Durham group has recently reported the discovery of TeV gamma ray emission from
PKS 2155-304 \cite{C1,C2}. The TeV  emission was detected in 1996 September and 1997 October/November, with 
the largest fluxes being measured in 1997 November. During 1997 November, we detected a record high  GeV gamma-ray
flux from PKS 2155-304 with CGRO/EGRET \cite{K1} and subsequently very high x-ray fluxes were measured
with BeppoSAX \cite{C3}. Here we report, for the first time, on the record x-ray fluxes measured with
the Rossi X-Ray Timing Explorer (RXTE) during the GeV/TeV outburst.

\section{X-Ray Observations}

In November 1997, after our detection of an extremely high GeV gamma-ray flux from
PKS 2155-304, we began a short series of target of opportunity observations with RXTE to
determine the x-ray properties of the source during the gamma-ray flare. Specifically, we made Proportional
Counter Array (PCA) and High Energy X-ray Timing Experiment (HEXTE) observations of nominal 2.5 ksec duration
on 20, 21, and 22 November 1997. Our analysis of those data indicate that the x-ray
flux on 20 and 21 November 1997 was $F(2-10\ $keV$)=2.3 \times 10^{-10}$ erg cm$^{-2}$s$^{-1}$.
By 17:00 UT on 22 November, when BeppoSAX measured its highest flux value \cite{C4}, the flux had slightly
dropped to $1.6 \times 10^{-10}$ erg cm$^{-2}$s$^{-1}$. The 20-21 November x-ray fluxes measured by RXTE 
are the highest ever observed in the 2-10 keV band for PKS 2155-304.

The x-ray spectral shapes we measured during the 1997 November flare show downward curvature consistent
with the idea that the synchrotron component is rolling off at keV energies.
If we fix the column depth at the galactic value n$_H=1.36 \times 10^{20}$cm$^{-2}$ \cite{L1},
then the 2.5-30.0 keV spectrum measured by the PCA on 20 November 1997 at 22:44-23:39 UT cannot be fit with a single
power law. However, a good fit is obtained if we use a broken power law 
with a low energy photon index $\alpha_{L}=2.51(\pm0.08)$, a high-energy photon index $\alpha_{H}=3.04(\pm0.03)$, and a
break energy $E_{b}=4.01(\pm0.22)$ keV. The 21 November measurements taken at 15:16-15:41 UT show a similar 
spectrum with $\alpha_{L}=2.72(\pm0.08)$, $\alpha_{H}=3.06(\pm0.04)$, and $E_{b}=4.33(\pm0.40)$\ keV. 
While yielding a slightly lower flux, the 22 November measurements taken at 17:00-17:33 UT are also
well fit by a broken power law but with parameters: $\alpha_H=2.98(\pm0.03)$, $\alpha_L=2.20(+0.30/-1.41)$
and a break energy $E_b=3.50(\pm0.45)$ keV.

To act as control observations, we examined PCA measurements of PKS 2155-304 made when either the
TeV or GeV gamma-ray fluxes were known to be low. During 30 December 1997-13 January 1998 we made
follow-up EGRET observations of the GeV emission from PKS 2155-304 and marginally detected GeV flux
at a level approximately a factor of four smaller than the November 1997 peak flux. Our
simultaneous PCA observations on 9-11 January 1998 measured 2-10 keV fluxes that had decreased
by a factor of seven from those observed on 20-21 November 1997.  Unlike the November 1997 spectra,
the 2.5-15.0 keV spectrum measured on 9 January 1998 at 2:59-3:40 UT can be acceptably fit
($\chi^{2}_{\nu}=0.95$ for 32 d.o.f.) using the galactic column depth and a single power law
having photon index $\alpha=2.83(\pm0.04)$. Chadwick et al. \cite{C1,C2} report detection of significant TeV flux from
PKS 2155-304 in September 1996, however the TeV flux apparently decreased and they were unable
to detect it in October or November 1996. While we do not have simultaneous TeV and x-ray observations,
the contemporaneous PCA observations made on 14 November 1996 show an x-ray flux which is a factor of
five smaller than those observed during the November 1997 TeV gamma-ray flare. Our observations are
therefore consistent with the pattern of correlated x-ray and gamma-ray flux outbursts observed
in the two well-studied TeV emitting XBLs, Mrk 421 and Mrk 501 \cite{B1}.

\begin{figure*}
\centerline{\epsfig{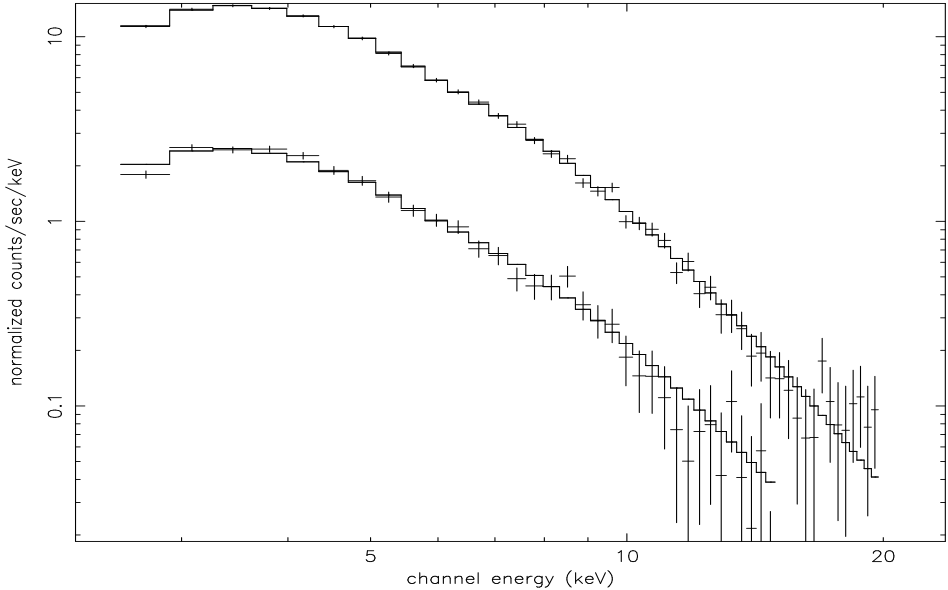}}
  \vspace{3cm}
\caption{PKS 2155-304 x-ray count spectra and folded photon models derived from RXTE/PCA measurements
taken on 20 November 1997 (top) and 9 January 1998 (bottom).}
\label{f1}
\end{figure*}

Measurements taken with the All-Sky Monitor (ASM) aboard the RXTE satellite also suggest a correlation
between elevated x-ray and gamma-ray emission. While substantially less sensitive than the PCA, the broad
field of view of the ASM provides much better temporal coverage and is sensitive enough
to detect major flaring activity from PKS 2155-304. Comparison of monthly ASM counting rates derived
by averaging over days when TeV observations were made with 5 months of TeV gamma-ray monthly counting 
rates indicates a positive correlation \cite{C1}. Our GeV gamma-ray measurements with EGRET suggest
that the correlation between gamma-ray and x-ray flux exists on even shorter timescales (see Figure 2). Subdivision
of the November 1997 EGRET observations indicates that the bulk of the $>$100 MeV emission was detected during
11-14 November. While the statistics are poor, measurements by the ASM hint at a strong x-ray flare
on 12-13 November simultaneous with the GeV flare and perhaps a second smaller flare on 19-20 November. Since the TeV gamma-ray 
and pointed x-ray observations did not begin until the 19th and 20th respectively, we suspect that they 
missed an even larger outburst on 12-13 November.

\begin{figure*}
\centerline{\epsfig{file=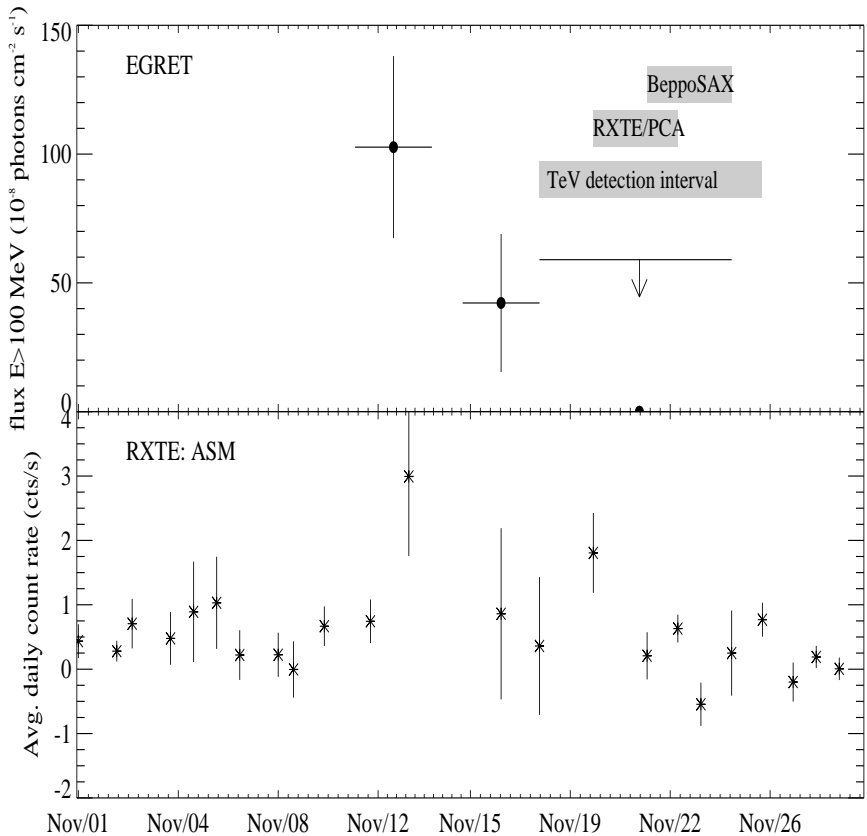,width=2.5in,height=3in,
bbllx=200pt, bblly=180pt,bburx=450pt,bbury=580pt,angle=90}}
  \vspace{3cm}
\caption{A comparison of the average daily count rate in the 2-10 keV  x-ray band measured by RXTE/ASM
with the gamma-ray flux measured above 100 MeV by CGRO/EGRET. Plotted are measurements of PKS 2155-304
 taken during November 1997. The shaded horizontal bars indicate the time intervals when TeV gamma-ray and
 pointed x-ray observations were made.}
\label{f2}
\end{figure*}

\section{Concluding Remarks}

The available PKS 2155-304 data show a correlation between the 2-10 keV x-ray outbursts and GeV/TeV
gamma-ray outbursts.  The ASM on RXTE has demonstrated the importance of all-sky x-ray monitoring as a trigger for
gamma-ray studies of XBLs and other blazars. The utility of this technique is currently limited by the sensitivity
and duty cycle of the ASM.  We expect that the launching of MOXE $-$ an all-sky x-ray monitor
which is a factor of four more sensitive than the ASM and has a duty cycle of nearly unity \cite{I1} $-$ and GLAST, the next generation
GeV gamma-ray telescope, in conjunction with the construction of the ground-based VERITAS array will initiate an exciting new era of
blazar study.

\end{document}